# The symmetry and light stuffing of $Ho_2Ti_2O_7$, $Er_2Ti_2O_7$ and $Yb_2Ti_2O_7$ characterized by synchrotron X-ray diffraction

Kristen Baroudi [a], Bruce D. Gaulin [b, c, d], Saul H. Lapidus [e] and R. J. Cava [a]

[a] Department of Chemistry, Princeton University, Princeton, NJ 08544 USA

[b] Department of Physics and Astronomy, McMaster University, Hamilton, ON L8S 4M1 Canada

[c] Brockhouse Institute for Materials Research, McMaster University, Hamilton, ON L8S 4M1, Canada

[d] Canadian Institute for Advanced Research, 180 Dundas Street W, Toronto, ON M5G 1Z8, Canada

[e] X-ray Science Division, Advanced Photon Source, Argonne National Laboratory, Argonne, IL 60439, USA

**Abstract**

The $Ho_2Ti_2O_7$, $Er_2Ti_2O_7$ and $Yb_2Ti_2O_7$ pyrochlores were studied by synchrotron X-ray diffraction to determine whether the (002) peak, forbidden in the pyrochlore space group *Fd-3m* but observed in single crystal neutron scattering measurements, is present due to a deviation of their pyrochlore structure from *Fd-3m* symmetry. Synchrotron diffraction measurements on precisely synthesized stoichiometric and non-stoichiometric powders and a crushed floating zone crystal of $Ho_2Ti_2O_7$ revealed that the (002) reflection is absent in all cases to a sensitivity of approximately one part in 30,000 of the strongest X-ray diffraction peak. This indicates to high sensitivity that the structural space group of these rare earth titanate pyrochlores is *Fd-3m*, and that thus the (002) peak observed in the neutron scattering experiments has a non-structural origin. The cell parameters and internal strain for lightly stuffed $Ho_{2+x}Ti_{2-x}O_7$ are also presented.

**Introduction**

The rare earth titanate pyrochlores, materials with formula $R_2Ti_2O_7$ (R = rare earth element), are archetype geometrically frustrated magnets. [1, 2] They exhibit a variety of magnetic states, including spin ice in $Ho_2Ti_2O_7$[3] and $Dy_2Ti_2O_7$[4], spin liquid behavior in $Yb_2Ti_2O_7$[5] and $Tb_2Ti_2O_7$[6] and antiferromagnetism and order by disorder in $Er_2Ti_2O_7$.[7, 8] The interpretation of these properties is critically dependent on the crystal structure and its symmetry, due to the influence of these factors on the magnetic interactions, and so it is important for the crystallographic space group of the materials to be well determined. The space group of the pyrochlores has long been reported as *Fd-3m* (#227) [9], but single crystal neutron scattering experiments sometimes reveal peaks in $R_2Ti_2O_7$ compounds, such as $Ho_2Ti_2O_7$[10], $Er_2Ti_2O_7$[11] and $Tb_2Ti_2O_7$[12, 13], that are forbidden by this space group, particularly the (002) reflection. If this diffraction peak is structural in origin, this would imply that the rare earth titanate structures actually crystallize in a subgroup or supergroup of *Fd-3m*. Neutron diffraction can contain harmonic Bragg contamination ($\lambda/n$) but this forbidden peak also appears in time of flight experiments, where such contamination is not expected. While this would seem to point to a structural origin, remnant magnetic fields or other non-structural effects, such as multiple scattering from systematically-allowed Bragg reflections, may also cause them to appear.

Here we use synchrotron powder X-ray diffraction data from beamline 11-BM at the Advanced Photon Source to study $Ho_2Ti_2O_7$, $Er_2Ti_2O_7$ and $Yb_2Ti_2O_7$ to determine whether there are peaks present that would indicate a space group that is different from *Fd-3m*. Synchrotron powder X-ray diffraction on this beamline has extremely high angular resolution, large dynamic range, is free from $\lambda/2$ harmonic Bragg contamination, is not sensitive to magnetic ordering, and does not operate under a magnetic field. Thus the observation of a *Fd-3m*-forbidden (002) peak at 11-BM would indicate that the space group of the rare earth titanate pyrochlores is not $Fd$-$3m$. Variations in the rare earth (R) to titanium ratio are possible in these compounds, leading to "stuffing' [14-17], and so several compounds with small differences in the R to Ti ratios were studied to determine whether the space group is sensitive to small deviations in stoichiometry. In no instance was the (002) reflection observed at room temperature in any of the materials. Thus the symmetry of the pyrochlores is $Fd$-$3m$. Because there has been no indication of a low temperature structural phase transition in rare earth titanate pyrochlores, which have been studied extensively by many characterization methods [2, 18, 19], the $Fd$-$3m$ symmetry must by inference be maintained to very low temperatures. Here we confirm this inference through performing a high-sensitivity synchrotron X-ray diffraction measurement on $Ho_2Ti_2O_7$ at 10 K. The (002) reflection is also not observed and thus the $Fd$-$3m$ space group is maintained to low temperature.

**Experimental**

Powders of $Ho_{1.98}Ti_{2.02}O_7$, $Ho_2Ti_2O_7$, $Ho_{2.02}Ti_{1.98}O_7$ (3 samples), $Ho_{2.04}Ti_{1.96}O_7$, $Ho_{2.06}Ti_{1.94}O_7$, $Er_2Ti_2O_7$, $Yb_{1.98}Ti_{2.02}O_7$, $Yb_2Ti_2O_7$ and $Yb_{2.02}Ti_{1.98}O_7$ were synthesized by mixing $Ho_2O_3$, $Er_2O_3$ or $Yb_2O_3$ powder with $TiO_2$ powder in a stoichiometric ratio in 3 gram batches, which are easily weighed out, to assure their stoichiometry. The powders were ground thoroughly in an agate mortar and pestle and then heated in air at 1400 °C for 48 hours. After heating, the samples were ground thoroughly and heated again at 1400 °C in air, with this cycle repeated several times. For $Ho_{1.98}Ti_{2.02}O_7$, $Ho_2Ti_2O_7$, $Ho_{2.02}Ti_{1.98}O_7$ (sample a), and $Er_2Ti_2O_7$ this was sufficient to form pure phase samples according to laboratory X-ray diffraction. Two of the samples of $Ho_{2.02}Ti_{1.98}O_7$ (samples b and c), $Ho_{2.04}Ti_{1.96}O_7$, $Ho_{2.06}Ti_{1.94}O_7$, $Yb_{1.98}Ti_{2.02}O_7$, $Yb_2Ti_2O_7$ and $Yb_{2.02}Ti_{1.98}O_7$, however, required additional heating to 1600 °C, which was performed in an argon-back-filled MRF Inc. furnace followed by quenching.[20] Heating at 1600 °C under argon caused the compounds to become slightly reduced (i.e. containing $Ti^{3+}$ at the ppm level), evidenced by a pale gray color, and so they were post annealed at 500 °C in air for 48 hours; after annealing the samples turned pink ($Ho_{2+x}Ti_{2-x}O_7$) or white ($Yb_{2+x}Ti_{2-x}O_7$) indicating that ppm levels of $Ti^{3+}$ were no longer present. These samples were then investigated by synchrotron powder X-ray diffraction at the Advanced Photon Source at Argonne National Laboratory, beam line 11-BM. A specimen from a single crystal of $Ho_2Ti_2O_7$ grown via the floating zone (FZ) method was ground into a powder and also investigated at 11-BM. Diffraction patterns were taken at 295 K with wavelengths 0.413804 Å, 0.413809 Å, 0.413852 Å, or 0.459012 Å. A 2θ range was selected so that all 12 detectors would pass over the region between 4 and 16 in 2θ to maximize the counting statistics at the positions of the (002) and (006) peaks forbidden in space group *Fd-3m*. A second diffraction pattern of the stoichiometric $Ho_2Ti_2O_7$ sample was taken at 10 K using a helium cryostat with somewhat reduced, but still extremely good, counting statistics. Reitveld refinements of the diffraction patterns were performed using the FullProf software suite.[21] During the refinement, the amounts of rare earth and Ti were held at the as-made stoichiometry of the sample and the oxygen content, which for the small degrees of stuffing studied here was in the range of 7 ± 0.03, was held at 7 per formula unit and also not refined.

The single crystal neutron diffraction measurements were performed on floating zone crystals grown from a stoichiometric melt following a procedure described elsewhere.[22]

**Results and Discussion**

Examples of single crystal neutron diffraction patterns that distinctly show very weak (002) reflections are presented in Figure 1. This reflection is prohibited in the pyrochlore space group *Fd-3m* due to the *d*-glide plane and its presence is the motivation for the current study. A magnetic (002) reflection is known to be a consequence of the application of a [110] magnetic field, polarizing the so-called alpha-chains of spins in single crystal pyrochlore magnets. This occurs at very low applied magnetic fields ($< 0.1$ T) in the classical spin ice state of $Ho_2Ti_2O_7$.[10] It is thus imaginable that it could occur in the remnant magnetic field of a superconducting magnet cryostat, which was the sample environment employed in several such studies. However, pyrochlore magnets with antiferromagnetic correlations, such as $Er_2Ti_2O_7$, which would be much harder to polarize, also display a weak (002) Bragg peak in low temperature time-of-flight neutron measurements.[11] An (002) Bragg peak could also arise due to multiple scattering of two structurally-allowed Bragg peaks, which would resemble a single scattering event Bragg peak at the systematically-absent (002) position. This effect is much stronger in single crystals than in powders. Therefore a systematic study of cubic pyrochlore magnets in powder form, and using synchrotron x-ray techniques, which are only sensitive to structural Bragg peaks, is very timely.

The synchrotron powder X-ray diffraction patterns of $Ho_{1.98}Ti_{2.02}O_7$, $Ho_2Ti_2O_7$, $Ho_{2.02}Ti_{1.98}O_7$ (a, b and c), $Ho_{2.04}Ti_{1.96}O_7$, $Ho_{2.06}Ti_{1.94}O_7$, ground FZ single crystal $Ho_2Ti_2O_7$, $Er_2Ti_2O_7$, $Yb_{1.98}Ti_{2.02}O_7$, $Yb_2Ti_2O_7$ and $Yb_{2.02}Ti_{1.98}O_7$ were quantitatively refined, and they each fit the known pyrochlore structure. An example of the excellent fit between the standard *Fd-3m* model for pyrochlores and the full synchrotron diffraction pattern is shown in Figure 2. The very high quality of the diffraction patterns, and the very high intensity of the strongest diffracted peaks, are apparent in the figure. The length of the unit cell *a*, the *x* coordinate of the oxygen at position 48f in the space group and the isotropic thermal parameters are the only variable structural parameters. The high resolution, high precision diffraction patterns showed that a systematic change in the size of the unit cell dimension can be detected when samples with small degrees of stuffing are tested, for example in the Ho and Yb-based samples, see Table 1. The diffraction pattern for each sample was well described by the $R_2Ti_2O_7$-type (R = rare earth) rare earth titanate pyrochlore structure, but the $Ho_{2.04}Ti_{1.96}O_7$ sample had a small (about 1%) impurity that fit a fluorite structure with a larger unit cell than the majority phase (likely related to $Ho_2TiO_5$)[23]. In addition, the $Yb_{2.02}Ti_{1.98}O_7$ sample and the ground $Ho_2Ti_2O_7$ FZ single crystal displayed subtly split peaks in the very high resolution patterns at high angle, indicating the presence of two $R_2Ti_2O_7$-type phases with extremely close compositions; the pattern from the crushed floating-zone grown $Ho_2Ti_2O_7$ single crystal will be discussed later.

Figure 3 shows the shift in the (222) peak with $x$ in $Ho_{2+x}Ti_{2-x}O_7$. There is a clear trend of higher Ho content leading to larger lattice parameters, which makes sense because $Ho^{3+}$ is larger than $Ti^{4+}$[24]. This figure also reveals the effect that sample preparation has on the rare earth titanate pyrochlores: three samples were prepared with stoichiometry $Ho_{2.02}Ti_{1.98}O_7$ and the two that were prepared in the same way (samples b and c) have nearly identical peak positions while the one prepared at lower temperature (sample a) is noticeably offset. The origin of the offset in these samples is not understood, but it is possible that subtle changes in the arrangement of the oxygen vacancies (there are 2 oxygen sites) is responsible. Interestingly, the pattern of the crushed single crystal overlays most closely on the pattern of $Ho_{2.06}Ti_{1.94}O_7$ (see Figure 3) suggesting that the stoichiometry of the floating zone grown crystal is Ho rich, with about 3% Ho stuffing on the Ti site. A similar effect was observed in $Yb_2Ti_2O_7$ crystals grown by the floating zone method.[15]

Figure 3 also shows a change in peak width as a function of $x$ in $Ho_{2+x}Ti_{2-x}O_7$. To quantify this observation the peaks were fit with a pseudo Voigt function and the full width half maximum (FWHM) was determined for each peak. In Figure 4 the FWHM of the (222) peak is plotted as a function of $x$ and it can be seen that the peak width increases with Ho stuffing. In the main panel of Figure 4, a dashed line is provided as a guide to the eye; the $x = -.02$ point is not considered, for reasons described further below.

The increase in FWHM of the peak with Ho stuffing implies that structural disorder (from Ho on the Ti site) may be introducing strain to the crystal structure. The inset of Figure 4 shows a Williamson-Hall plot for $Ho_2Ti_2O_7$ and $Ho_{2.06}Ti_{1.94}O_7$. This plot takes advantage of the fact that peak broadening due to crystallite size and strain have different dependencies on Bragg angle: the former has a $1/\cos\theta$ dependence while the latter has a $\tan\theta$ dependence.[25, 26] The FWHM was corrected for instrumental broadening using the equation

$$\text{FWHM} = \left[\left(\text{FWHM}^2_{measured}\right) - \left(\text{FWHM}^2_{instrumental}\right)\right]^{\frac{1}{2}} \quad (1)$$

The equation for peak broadening due to both effects is

$$\text{FWHM} = \left(\frac{k\lambda}{D\cos\theta}\right) + 4\varepsilon \tan\theta \quad (2)$$

where k is the Scherrer constant, $\lambda$ is the wavelength, D is the crystallite volume and $\varepsilon$ is the weighted average strain. This rearranges to

$$\text{FWHM}\cos\theta = \left(\frac{k\lambda}{D}\right) + 4\varepsilon \sin\theta \quad (3)$$

which is the Williamson Hall equation. When FWHM $\cos\theta$ is plotted against $4\sin\theta$ the slope of the line is equal to the strain $\varepsilon$ in the crystal structure. This is the case for the stuffed pyrochlores studied here – thus the observed peak broadening is due to strain, not crystallite size. A comparison of the two compounds plotted in the inset of Figure 4 shows that $Ho_2Ti_2O_7$ displays a flat line, while $Ho_{2.06}Ti_{1.94}O_7$ displays straight line behavior with a significant slope. This indicates that there is little or no detectable strain in $Ho_2Ti_2O_7$ but that the Ho stuffing in $Ho_{2.06}Ti_{1.94}O_7$ leads to strain in the crystal structure; $\varepsilon$ is approximately equal to 2.

Figure 5 shows the synchrotron diffraction pattern of the $Ho_2Ti_2O_7$ ground FZ single crystal. The pattern is a good match for a pyrochlore structure with a slightly larger unit cell than the stoichiometric $Ho_2Ti_2O_7$ powder, in addition, the high angle peaks show very fine splitting (see the inset in Figure 5). The splitting can be fit by modeling two very similar pyrochlore phases with slightly different unit cell dimensions. The origin of the two very similar phases is likely to be a very small variation in stoichiometry of the single crystal; we can estimate this variation from the data shown in Figure 5 as 0.026 in the Ho content.

Figure 6a shows the variation of unit cell length (normalized to the stoichiometric $R_2Ti_2O_7$ composition) as a function of rare earth stuffing in $Ho_{2+x}Ti_{2-x}O_7$ and $Yb_{2+x}Ti_{2-x}O_7$ near $x = 0$. There is a trend of increasing unit cell length with increasing rare earth content, which is to be expected due to the larger size of $Ho^{3+}$ and $Yb^{3+}$ compared to $Ti^{4+}$. The unit cell length of $Ho_{2+x}Ti_{2-x}O_7$ responds more strongly to doping than $Yb_{2+x}Ti_{2-x}O_7$ based on the larger slope of the fit to the $Ho_{2+x}Ti_{2-x}O_7$ data. The difference between the atomic radii of $Ho^{3+}$ and $Yb^{3+}$ (0.901 Å and 0.868 Å, respectively)[24] does not appear to completely account for this trend. This plot thus implies, along with the fit of the "negative stuffing" point at $x = -0.02$ on the trend displayed by the $x = 0$ and $x = 0.02$ points, that in $Yb_{2+x}Ti_{2-x}O_7$ there is mixing of Yb and Ti on the A and B sites near the stoichiometric composition due to the more similar ionic radii of $Yb^{3+}$ and $Ti^{4+}$. In contrast, for $Ho_{2+x}Ti_{2-x}O_7$, the $x = -0.02$ data falls off the trend displayed for the positive stuffing data. This is also the case for the data shown in Figure 4 and the data shown in figure 6b (described below). The clear inference to be drawn from all three kinds of data analysis (i.e. Figure 4, Figure 6a and Figure 6b) is that "negative stuffing" (i.e. titanium on the rare earth site or vacancies on the rare earth site) in the $Ho_2Ti_2O_7$ pyrochlore is extremely limited; from the linear extrapolation of all three trends, the limit of negative stuffing is between $x = -0.003$ and $x = -0.005$ in $Ho_{2+x}Ti_{2-x}O_7$.

The oxygen atom bonded to the ions in the B-site octahedron (which is occupied by only Ti in the ideal case) is in position 48f and has one variable positional coordinate $x$ that is a reflection of the size of the ions in the octahedron. This coordinate is defined to high precision in the synchrotron diffraction refinements. The refinements show that the B - O distance in the B-site octahedron increases with stuffing concentration $x$ in $Ho_{2+x}Ti_{2-x}O_7$ as larger $Ho^{3+}$ ions mix in with the $Ti^{4+}$ present in the ideally stoichiometric material (see figure 6b). This increase corresponds roughly with that of the unit cell length and can be compared to the plot in figure 6a showing the normalized unit cell parameter $a/a_0$ versus the change in $x$.

After confirming that each compound has the pyrochlore structure, the synchrotron diffraction patterns can then be studied much more closely to look for signs of the (002) peak observed in the single crystal neutron scattering experiments. The region of the diffraction patterns where the forbidden (002) peak would be observed if present is shown in Figure 7; the arrows mark the expected location of the peak in each case, and the sensitivity of the diffracted pattern relative to the intensity of the strongest diffraction line is indicated by a vertical bracket. In these diffraction patterns it is clear that there are no peaks at that position in any of the samples down to the limit of detection, which is $\sim 10^{-4}$ times smaller than the intensity of the highest peak in the diffraction patterns (The very small extra peaks in $Er_2Ti_2O_7$ and $Ho_2Ti_2O_7$ at other positions are attributed to the presence of extremely small amounts of impurity phase). The absence of this (002) peak in the diffraction patterns of the studied rare earth titanate pyrochlores taken using very high intensity X-rays is strong evidence that the (002) peak is not structural in origin. The peak is also not present in any of the samples deliberately prepared with different levels of stuffing, and so it cannot be attributed to changes in symmetry due to small stoichiometry differences.

The synchrotron diffraction pattern of $Ho_2Ti_2O_7$ taken at 10 K was also examined to determine whether the (002) peak is present. The region where it would appear is shown in Figure 8 with the position of the absent peak marked by an arrow. This pattern shows that at 10 K $Ho_2Ti_2O_7$ has not undergone a structural distortion that would generate the (002) peak. The inset shows the (222) peak for the 295 K and 10 K patterns. The 10 K peak is shifted to higher Q, indicating that the unit cell has become smaller on cooling, which is expected. There is no splitting or broadening of the peak, indicating that the cubic dimensionality is very well maintained to low temperature. Thus, in agreement with physical property measurements that have shown no structural phase transitions in the rare earth titanate pyrochlores on cooling below ambient temperature, the high resolution synchrotron X-ray diffraction study on $Ho_2Ti_2O_7$ at low temperature shows no change from the ideal pyrochlore structure; the room temperature symmetry is maintained to low temperatures.

**Conclusions**

The high precision and accuracy of the synchrotron diffraction patterns allows the observation that increasing the Ho content in $Ho_{2+x}Ti_{2-x}O_7$ for very small degrees of stuffing near the stoichiometric $x = 0$ value leads to a detectable increase in the length of the unit cell and the average metal-oxygen bond length in the octahedral position, consistent with the stuffing model for rare earth rich titanate pyrochlores. A powdered single crystal of nominally stoichiometric $Ho_2Ti_2O_7$ grown by the floating zone method is shown to be slightly Ho-rich and further shows slight peak splitting at high angle in the very high resolution synchrotron X-ray characterization, indicating the presence of two pyrochlores with very similar compositions, which we estimate differ by $\Delta x = 0.026$ in the formula $Ho_{2+x}Ti_{2-x}O_7$.

The (002) peak is observed in some single crystal neutron diffraction experiments on the rare earth titanates $Ho_2Ti_2O_7$, $Yb_2Ti_2O_7$ and $Er_2Ti_2O_7$ despite the fact that it is forbidden in the *Fd-3m* space group. The present high sensitivity synchrotron X-ray diffraction experiments show that the (002) reflection is absent in these compounds and in stuffed pyrochlores with slight variations in stoichiometry. This indicates that the presence of the (002) reflections in neutron diffraction experiments is not due to a deviation of the rare earth titanate space group from *Fd-3m*. Given the sum of the evidence, we speculate that the most likely origin of this peak is multiple scattering from strong allowed Bragg reflections.

**Acknowledgements**

The authors would like to thank the staff at Argonne National Laboratory Beamline 11-BM for their help in designing the high sensitivity experiments. The research at Princeton was supported by the DOE through the IQM at Johns Hopkins University, grant DE-FG02-98-ER46544. The research at McMaster University was supported by NSERC of Canada. Use of the Advanced Photon Source and Argonne National Laboratory was supported by the U. S. Department of Energy, Office of Science, Office of Basic Energy Sciences, under Contract No. DE-AC02-06CH11357. Discussions with Michel Gingras, Satya Kushwaha, K.A. Ross, J. Gaudet, J.P. Clancy, and J.P. Ruff are gratefully acknowledged.

**Figure Captions**

Figure 1: (Color online) Elastic (-0.1 meV < E < 0.1 meV) neutron diffraction maps within the HHL plane of reciprocal space taken on $Er_2Ti_2O_7$, $Ho_2Ti_2O_7$ and $Yb_2Ti_2O_7$ single crystals at low temperatures (~ 0.1 K) using the DCS spectrometer at NIST. The weak (002) Bragg reflection is highlighted.

Figure 2: (Color online) Synchrotron X-ray diffraction pattern at 295 K of a polycrystalline sample with nominal composition $Ho_2Ti_2O_7$. Red squares are observed intensity, the black line is calculated intensity, the blue line is the difference calculation and *hkl* reflections are green dashes. Inset: a blown up view of the (884)/(12 0 0) reflection. Red squares are observed intensity, the black line is calculated intensity and *hkl* reflections are green dashes.

Figure 3: (Color online) The (222) peak is shown for each pattern. The intensity of each peak is normalized to 1 and is plotted against Q to compare across different X-ray wavelengths. The squares are data points and the lines are a guide to the eye.

Figure 4: (Color online) The full width half maximum (FWHM) of the (222) peak of $Ho_{2+x}Ti_{2-x}O_7$ is plotted as a function of *x*. The error bars are smaller than the points. The dashed line is a guide to the eye. The inset shows a Williamson-Hall plot of $Ho_2Ti_2O_7$ and $Ho_{2.06}Ti_{1.94}O_7$ with linear fits; this plot indicates that strain is present in the stuffed pyrochlores but not in the stoichiometric pyrochlore. The equation of the fit to $Ho_2Ti_2O_7$ is y = 0.58 - 0.09x and the equation of the fit to $Ho_{2.06}Ti_{1.94}O_7$ is y = 0.33 + 1.89x.

Figure 5: (Color online) Synchrotron X-ray diffraction pattern of a ground FZ single crystal with nominal composition $Ho_2Ti_2O_7$. Red squares are observed intensity, the black line is calculated intensity, the blue line is the difference calculation and *hkl* reflections are green dashes. The two rows of *hkl* reflections correspond to two different phases. Inset: a blown up view of the (884)/(12 0 0) reflection showing peak splitting. Red squares are observed intensity, the black line is calculated intensity and *hkl* reflections are green dashes.

Figure 6: (Color online) a) The unit cell length *a* normalized to the unit cell length for the undoped phase $a_0$ as a function of doping amount *x*. $Ho_{2+x}Ti_{2-x}O_7$ is represented by black squares and $Yb_{2+x}Ti_{2-x}O_7$ is represented by red triangles. Error bars are smaller than the symbols. The lines are linear fits to the data. b) The average B site – O bond length in angstroms as a function of doping content *x* for $Ho_{2+x}Ti_{2-x}O_7$. Data points are black squares bracketed by error bars (for most data points the error is smaller than the symbol). The line is a linear fit to the data.

Figure 7: (Color online) Zoomed in views of synchrotron diffraction patterns at 295 K of selected compounds near the (002) peak. The y axis is intensity in arbitrary units while the x axis is Q ($Å^{-1}$). Each pattern is labeled by the chemical formula directly above it. The vertical arrows above each pattern mark where the (002) peak would appear if it were present. The patterns have been normalized and offset in intensity. A scale bar in the upper right corner of the panel shows

the height of a peak that would be 0.0001 times the maximum intensity for a given pattern. The peaks around $Q = 1.29$ Å$^{-1}$ in $Er_2Ti_2O_7$ and $Ho_2Ti_2O_7$ are from small amounts of unknown impurities.

Figure 8: (Color online) Detailed view of the critical region of the 10 K synchrotron diffraction pattern of $Ho_2Ti_2O_7$. The arrow points to where the (002) peak would be if it were present. The scale bar on the left shows the height of a peak that would be 0.0001 times the maximum intensity for the strongest peak in the pattern. The inset shows a comparison between the normalized (222) peak for the 295 K synchrotron diffraction pattern (green squares) of $Ho_2Ti_2O_7$ and the same peak for the 10 K pattern (brown squares). The lines are a guide to the eye.

Table 1: $Ho_{2.02}Ti_{1.98}O_7$ (a) was heated to 1400 °C in air, (b) and (c) were heated to 1600 °C in argon. $\chi^2$ is the goodness of fit of the calculated diffraction pattern.

Table 1

| Compound | a (Å) | 48f x,0,0 | Avg B site-O bond (Å) | $\chi^2$ | # phases |
|---|---|---|---|---|---|
| $Ho_{1.98}Ti_{2.02}O_7$ | 10.102200(1) | 0.32836(1) | 1.95342(4) | 3.271 | 1 |
| $Ho_2Ti_2O_7$ | 10.103421(1) | 0.32855(1) | 1.95443(7) | 3.847 | 1 |
| $Ho_{2.02}Ti_{1.98}O_7$ (a) | 10.106468(1) | 0.32909(3) | 1.9573(1) | 4.888 | 1 |
| $Ho_{2.02}Ti_{1.98}O_7$ (b) | 10.108794(1) | 0.32893(1) | 1.95711(4) | 1.608 | 1 |
| $Ho_{2.02}Ti_{1.98}O_7$ (c) | 10.109204(1) | 0.32876(1) | 1.95640(6) | 2.158 | 1 |
| $Ho_{2.04}Ti_{1.96}O_7$ | 10.111421(1) | 0.32958(1) | 1.96025(7) | 1.9602 | 2 |
| $Ho_{2.06}Ti_{1.94}O_7$ | 10.119061(1) | 0.32981(1) | 1.96267(8) | 3.334 | 1 |
| $Ho_2Ti_2O_7$ (FZ phase 1) | 10.121589(1) | 0.32937(1) | - | 3.945 | 2 |
| $Ho_2Ti_2O_7$ (FZ phase 2) | 10.118398(1) | 0.32937(1) | - | - | - |
| $Yb_{1.98}Ti_{2.02}O_7$ | 10.033221(1) | 0.33264(1) | 1.95790(8) | 1.997 | 1 |
| $Yb_2Ti_2O7$ | 10.034243(1) | 0.33176(4) | 1.9544(3) | 2.810 | 1 |
| $Yb_{2.02}Ti_{1.98}O_7$ | 10.035555(1) | 0.32774(2) | 1.9380(1) | 2.323 | 2 |
| $Er_2Ti_2O_7$ | 10.077044(1) | 0.32996(1) | 1.95513(4) | 3.224 | 1 |

Figure 1

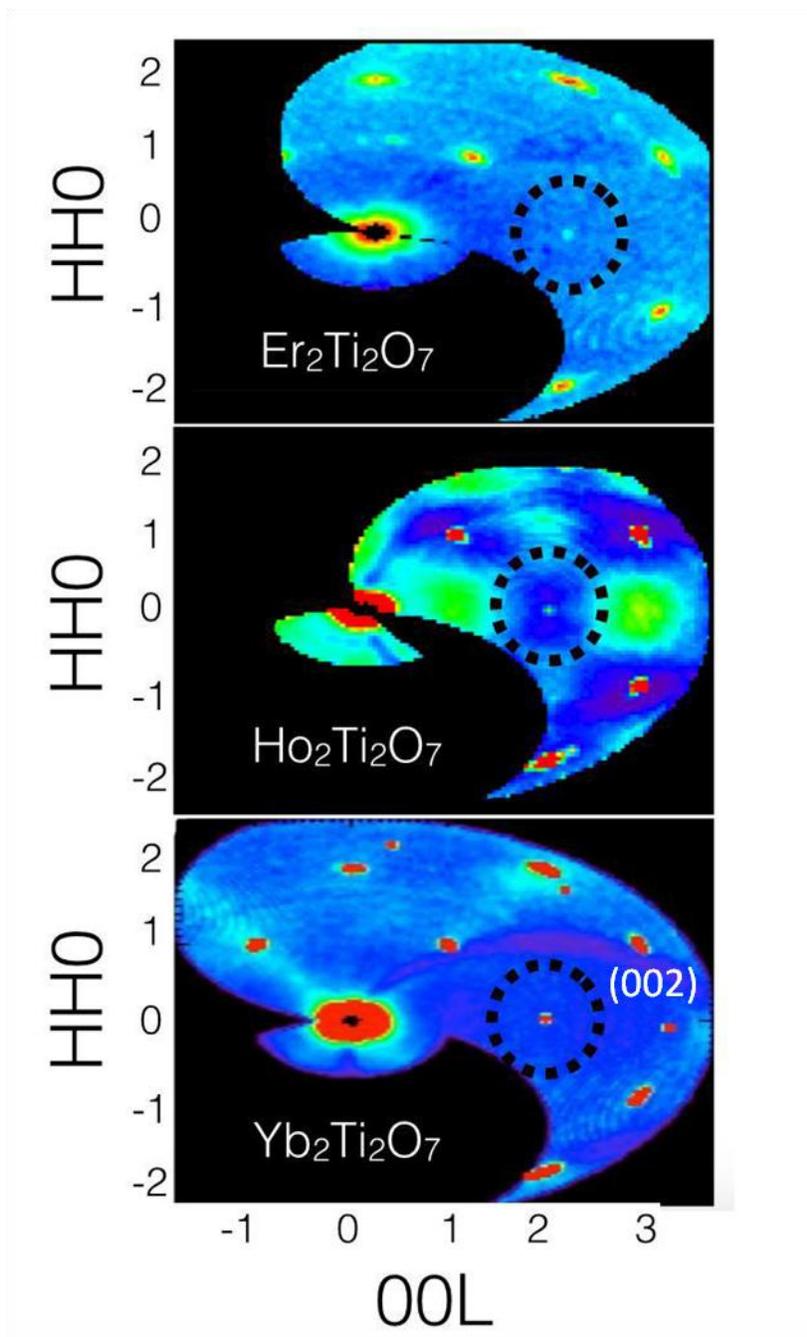

Figure 2

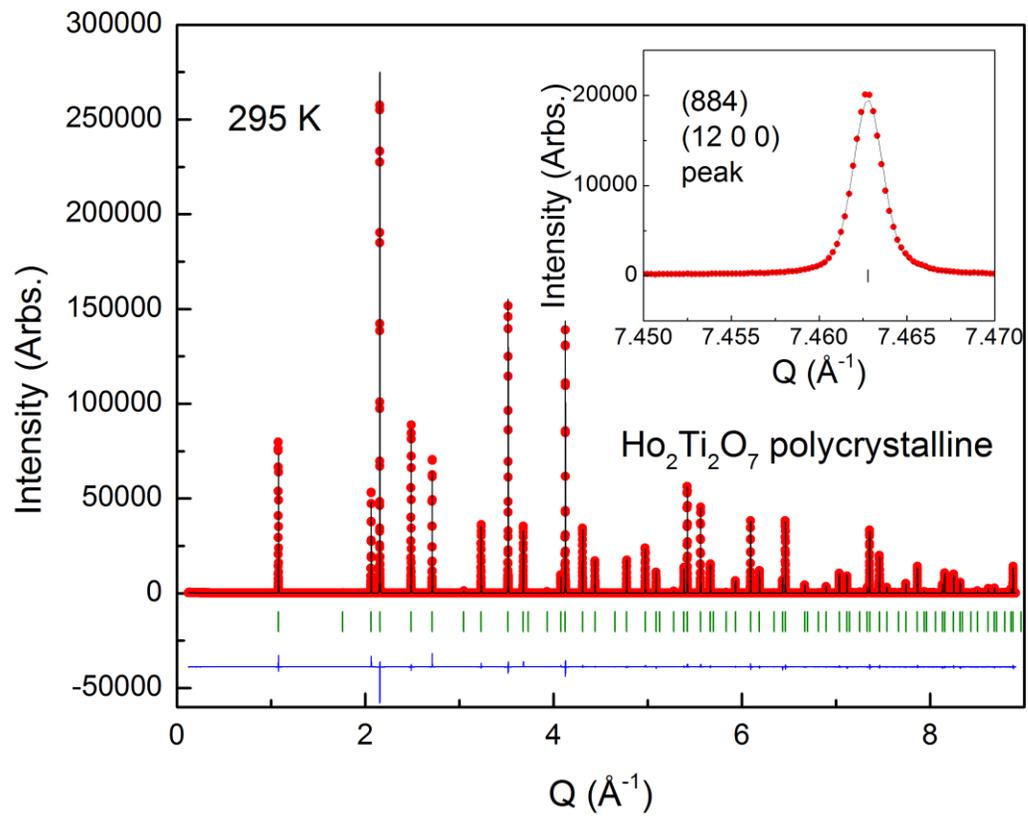

Figure 3

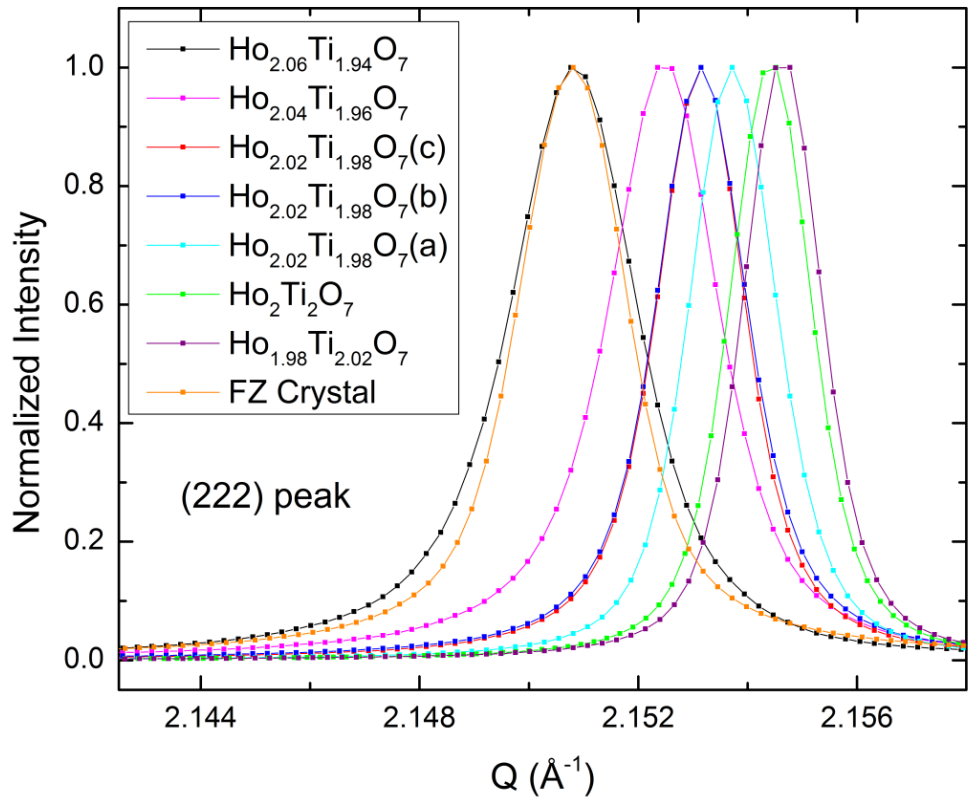

Figure 4

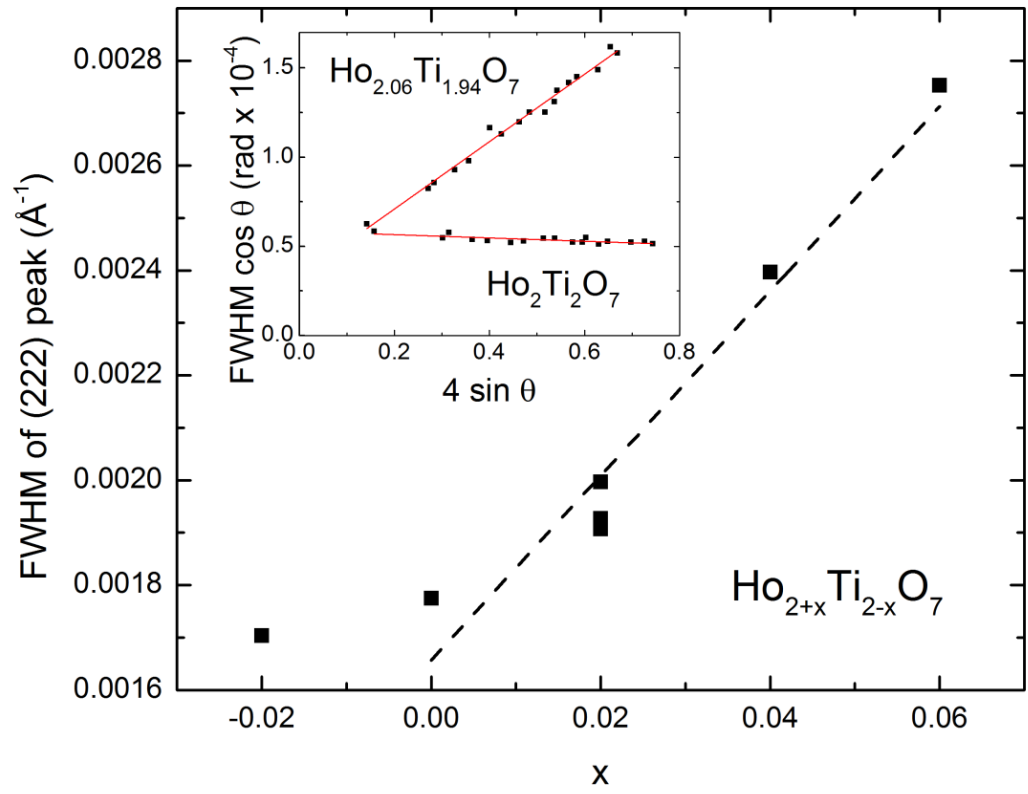

Figure 5

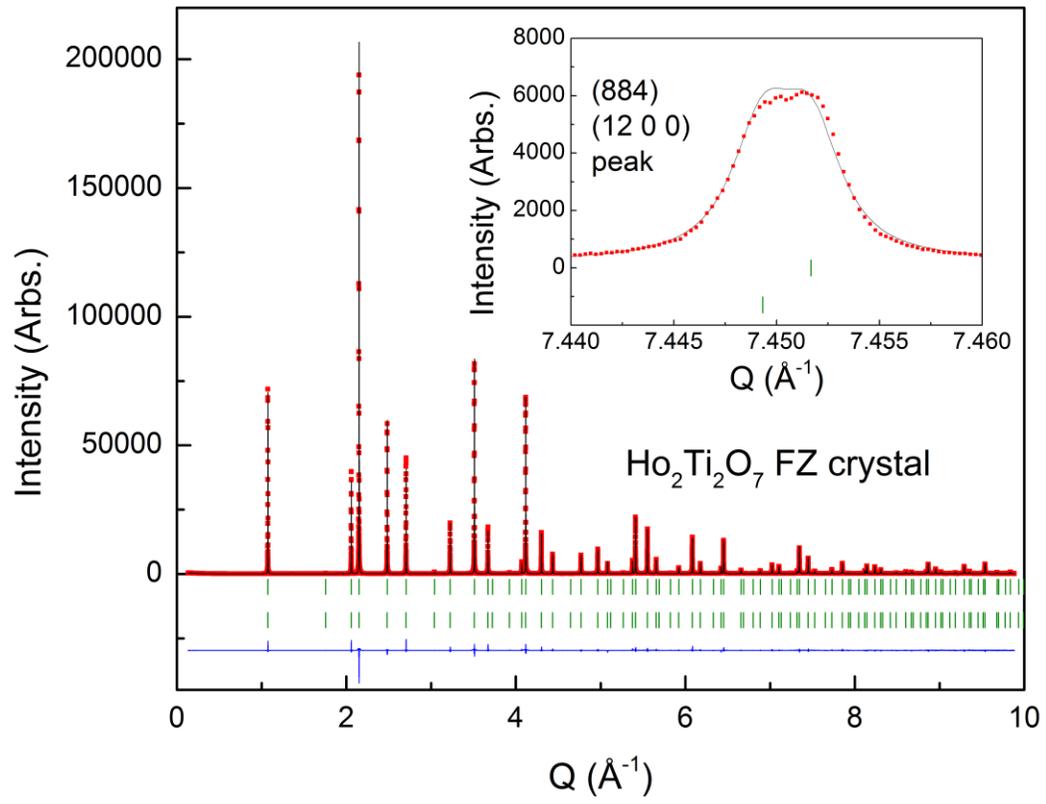

Figure 6

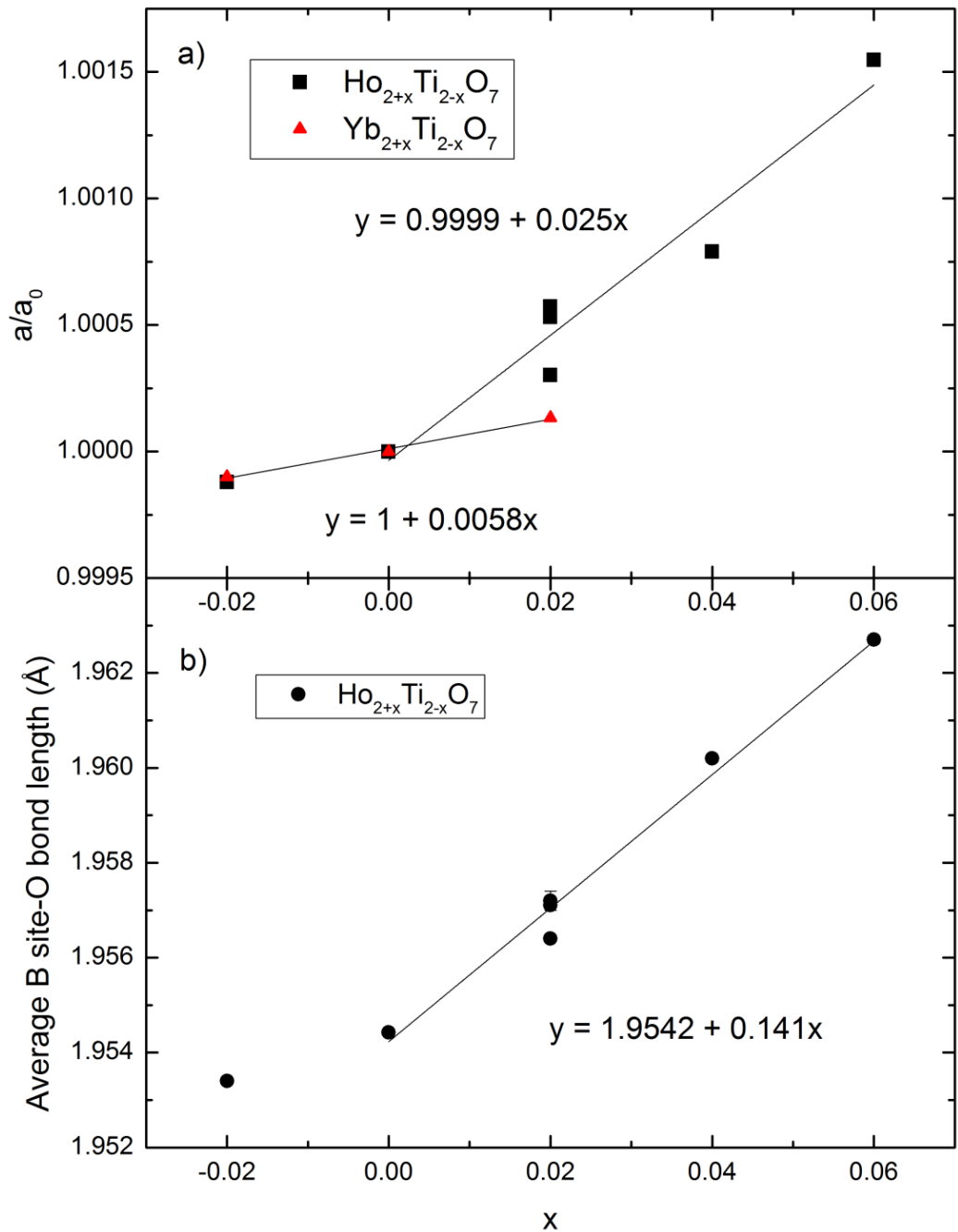

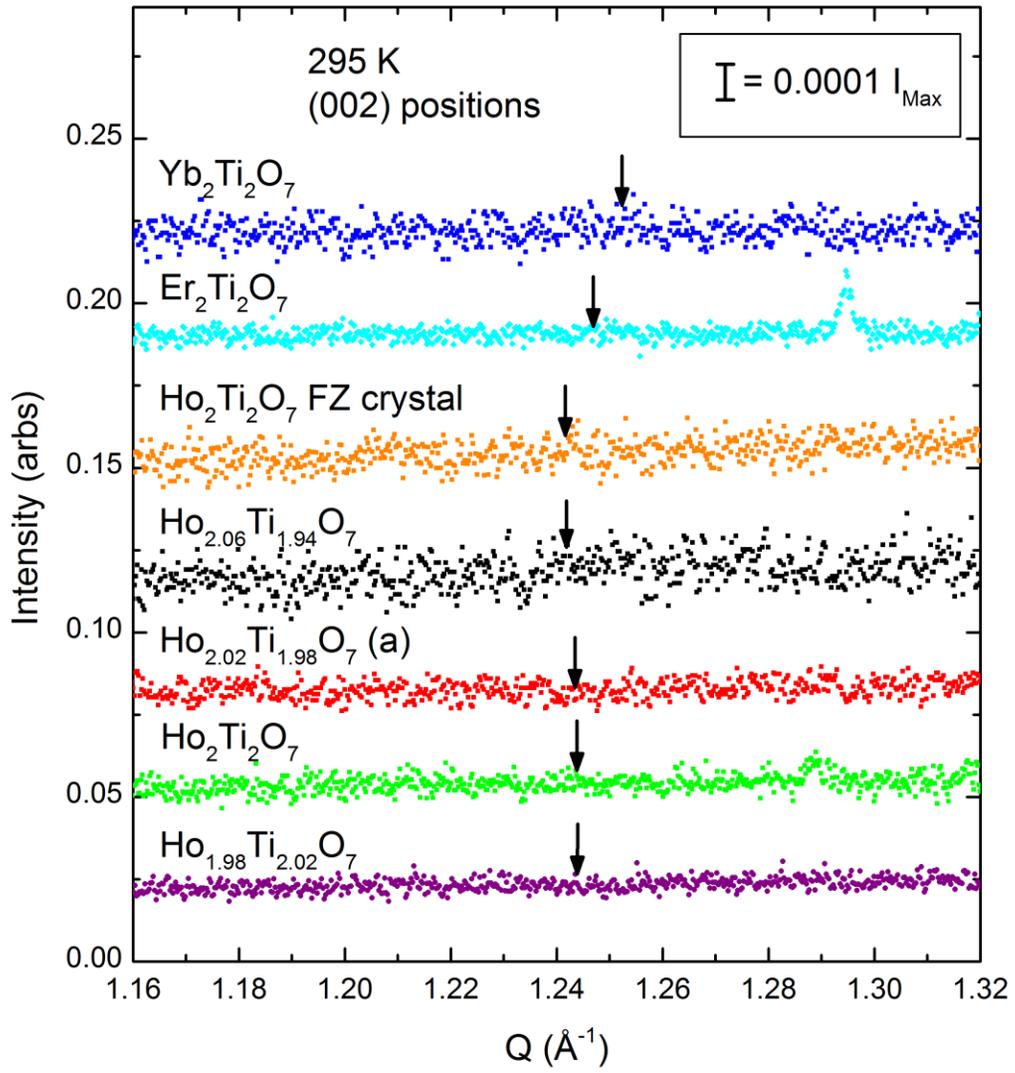

Figure 8

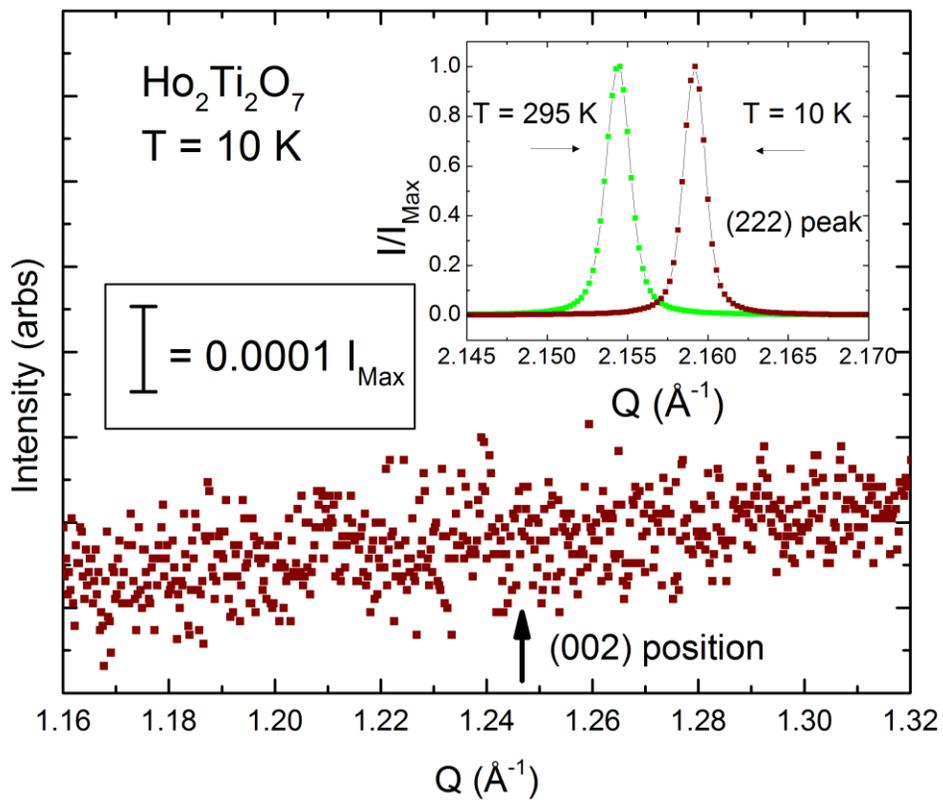